\documentstyle[12pt,aasms4]{article}
 
\slugcomment{September 3, 1997 -- to appear in the Astrophysical Journal}
 
\begin{document}
 
\title{Redshifted Neutral Hydrogen 21cm Absorption toward ~~~~~~~~~~~~
Red Quasars}
 
\author{C.L. Carilli}
\affil{NRAO, P.O. Box O, Socorro, NM, 87801, USA \\}
\authoremail{ccarilli@nrao.edu}
\author{Karl M. Menten}
\affil{Max Planck Institut f\"ur Radioastronomie, Auf dem Hugel 69,
D-53121, Bonn, Germany \\}
\author{Mark J. Reid}
\affil{SAO, 60 Garden St., Cambridge, MA, 02138, USA \\}
\author{M.P. Rupen and Min Su Yun}
\affil{NRAO, P.O. Box O, Socorro, NM, 87801, USA \\}  
 
\begin{abstract}

We have searched for redshifted neutral hydrogen 21cm absorption toward
sources from the Stickel et al. (1996a) `red quasar' sub-sample.
The red quasar sub-sample is taken from 
the 1 Jy sample of flat spectrum radio sources, and is comprised 
of the 15 sources which are 
undetected on the POSS.
Five of these red quasars have been searched for redshifted HI 21cm absorption
to optical depth levels of a few percent, and four show
strong absorption, with neutral hydrogen column densities between  
4 and 80$\times$10$^{18}$$\times({{T_s}\over{f}})$ cm$^{-2}$.
This 80$\%$ success rate for the red quasars
compares to the much lower success rate of only 11$\%$ 
for detecting HI 21cm absorption 
associated with optically selected Mg II absorption line systems (Briggs
and Wolfe 1983, Lane et al. 1997).
The large neutral hydrogen column densities seen 
toward the Stickel et al.  red quasars provide circumstantial 
evidence supporting the dust reddening hypothesis, as opposed to 
an intrinsically red spectrum for the AGN emission mechanism. 
The lower limits to rest frame values of A$_V$ are between 2 and 7, leading
to lower limits to the  spin temperatures for the neutral hydrogen
between 50 K and 1000 K, assuming a  Galactic dust-to-gas ratio.
 
We consider the question of biases in
optically selected samples of quasars due to dust obscuration. 
Overall, the data on the red quasar sub-sample
support the models  of Fall and Pei (1993)
for dust obscuration by damped Ly$\alpha$ absorption line systems,
and suggest that:
(i) there may be a significant, but not dominant, population of 
quasars missing from optically selected samples due to
dust obscuration, perhaps as high as 20$\%$ at the POSS limit
for an optical sample 
with a redshift distribution similar to the 1 Jy, flat spectrum quasar
sample, and (ii) optically
selected samples may miss about half the high column density
quasar absorption line systems.

The redshifted HI 21cm absorption line
detections presented herein are toward the sources:
0108+388 at z = 0.6685, 0500+019 at z = 0.5846, and 1504+377
at z = 0.6733. No absorption is seen 
toward 2149+056 at z = 0.740 at a level below that seen for the three 
detections, although there is some uncertainty in this case
as to the expected line redshift.
In some systems the absorbing gas is in the vicinity of the AGN,
either circumnuclear material or 
material in the general ISM of the AGN's host galaxy, as is probably the case
for 0108+388 and 1504+377, and in other systems the absorption is
by gas associated with galaxies cosmologically distributed along
the line of sight to the quasar, as may be the case for 0500+019.
The WSRT spectrum of 1504+377
confirms the lack of HI 21cm absorption associated 
with the narrow molecular absorption line system at z = 0.67150.

\end{abstract}
 

\section {Introduction}

K\"uhr et al. (1981) have compiled a complete,
all-sky sample of 299 radio sources
with flux densities of 1 Jy or more at 5 GHz, and 
with flat radio spectral indices  between 5 GHz and 2.7 GHz 
($\alpha_5^{2.7}$ $\ge$ $-0.5$, where spectral index is defined as:~
S$_{\nu}$ $\propto$ $\nu^{\alpha}$). In the process of optical 
identification of all sources at declinations greater than $-$20$^{o}$
in this sample, Stickel et al. (1996a) have found that 15 
sources are undetected on the Palomar Observatory
Sky Survey (POSS) plates.
They designate these 1 Jy, flat spectrum radio sources with faint
optical counter-parts as 
`red quasars'.  Subsequent deep imaging  in the optical and near IR by
Stickel et al. (1996a) shows that these red quasars typically have
faint, extended, galaxy-like emission in the 
optical, with little or no evidence for a bright point source,
while near IR images reveal a dominant point source in most cases.
Moreover, 13 of the 15 sources are  variable in the near IR,
indicating emission from an active galactic nucleus (AGN), as opposed
to a starburst. Stickel et al. (1996a) consider two physical models  
for the red colors of these sources:
an intrinsically red spectrum for the AGN emission mechanism, perhaps
due to a high energy cut-off in the relativistic electron population
(assuming the emission is synchrotron radiation; Bregman et al. 1981),
or reddening of the spectrum by intervening dust. 

Radio selected quasar samples, such as the 1 Jy, flat spectrum source
sample, are important because they avoid possible biases
introduced in optically selected quasar samples by dust obscuration along
the line of sight. Such obscuration may have a
significant impact on 
the numbers of quasars seen at high redshift in optical samples, and
on the number of high column density 
absorption line systems seen toward optically selected 
quasars (Heisler and Ostriker 1988,
Fall and Pei 1993, Webster et al. 1995, Shaver et al. 1996).

In this paper we present a search for redshifted neutral hydrogen 21cm
absorption toward sources  from the Stickel et al. (1996a)  red quasar
sub-sample using the new UHF system at the Westerbork Synthesis Radio
Telescope (WSRT). These observations are intended to 
address a number of issues. First is the 
question of whether some AGNs have intrinsically red,
rather than dust-reddened, emission spectra. 
The detection of a large column of neutral hydrogen would provide at
least circumstantial evidence in favor of dust-reddening ({\sl cf.}
Carilli, Perlman, and Stocke 1992).
Second is the question of the fraction of quasars missing from optical
samples due to dust obscuration, as well as the fraction of missing
high column density quasar absorption line systems.
And third is to enlarge the
sample of dusty, high column density absorption line systems 
in order to study the dense ISM in 
galaxies at significant look-back times.

\section{Observations}

The WSRT  is in the process of a major up-grade of the receivers
and correlator (A.-J. Boonstra 1996). One goal of this up-grade is 
to provide almost
continuous frequency coverage from 250 MHz to  1420 MHz. 
This system offers a new window for studies of redshifted neutral
hydrogen using an interferometer. Interferometers offer significant
advantages over single dishes when searching for faint, broad
lines at low frequency, including the  natural mitigation of
terrestrial interference due to fringe rotation, more accurate
continuum flux density determination, and more linear spectral baselines.

We have used the UHF-high part of this system (750 MHz to 1000 MHz) to
search for redshifted HI 21cm absorption toward sources from the Stickel
et al. (1996a) red quasar sub-sample. The redshifts used in the search are
taken from Table 1 of Stickel et al., and are 
discussed in more detail in section 3. The HI 21cm line is
redshifted into the UHF high band of the WSRT for six of the nine sources 
with known redshifts in Table 1 of Stickel et al. (1996a).
Redshifted HI 21cm absorption has already been detected against 
two of the six sources (0218+357 and 1504+377) using the 140 ft
telescope of the National Radio Astronomy Observatory (NRAO)\footnote{
The National Radio Astronomy (NRAO) is a facility
of the National Science Foundation, operated under cooperative
agreement by Associated Universities, Inc.} 
in Green Bank, WV (Carilli et al. 1993, Carilli et al. 1997).
However, the spectrum of 1504+377
was marred by terrestrial interference, and so we re-observed this 
source with the interferometer.

The sources observed with the WSRT are listed in Table
1, along with observing dates in column 2. Column 3 lists the 
frequency for the observing band-centers (corrected to 
a heliocentric rest-frame), and column 4 lists
the source continuum flux densities at the observed frequency, as determined
from off-line channels. The error bars reflect a
10$\%$ uncertainty in absolute calibration.
Columns 5, 6, and 7 give the on-source integration time, total bandwidth,
and number of frequency channels. Columns 8 and 9 give the velocity
resolution and the root mean square (rms) noise in the final spectra.
Note that all of these sources are phase calibrators for the NRAO Very
Large Array, and accurate positions can be found in Taylor and Perley
(1996). We use the B1950 designation for the sources, in keeping with
previous papers.

Both linear polarizations were received for each of the 40 baselines
connecting the 10 fixed and 4 move-able Westerbork antennas, although
for some observations one or two of the antennas were inoperable.
The data were gain, phase, and bandpass
calibrated using the Astronomical Image Processing System (AIPS)
at the National Radio Astronomy Observatory in Socorro, NM.
The quasars  3C 286, 3C 48, and 3C 147 were observed for absolute flux
calibration, according to the scale of Baars et al. (1977), and for
initial antenna-based phase calibration. Self-calibration on the
target continuum sources was then
employed for further refinement of the antenna-based phase solutions.
The  absolute gain calibration 
accuracy is about 10$\%$, as estimated by applying
calibration from one standard calibrator to another. 
Data were inspected for interference and other problems, and bad data
were removed from the visibility data-set. In the case of 0202+149, 
strong terrestrial interference (TV channel 48) 
was present in the observing band close to the
line frequency, precluding observation of this source by the WSRT.

The calibrated and edited data were Fourier transformed to 
produce images for each frequency channel, and these channel images were 
assembled into an image `cube'. The AIPS program `IMLIN' was then used
to subtract the continuum emission in the field by fitting a linear
baseline to off-line channels at each pixel in the image cube.
The subsequent continuum-subtracted  image cubes were then 
deconvolved using the Clark CLEAN algorithm as implemented
in the AIPS task `APCLN' (Clark 1980). In each case
the CLEANing was terminated when residual components equaled twice the
rms noise per channel. The Gaussian restoring CLEAN beams for each source
are listed in the figure captions. Radio spectra were then extracted at
the position of the continuum source, and converted
to optical depth using the continuum source flux density. Finally, these
optical depth spectra were  analyzed for 
integrated optical depth and velocity structure by fitting Gaussians. 
All redshifts quoted have been corrected to the heliocentric 
reference frame. Three of the  four 
target sources were unresolved by the WSRT, with the exception of
0108+388, as discussed in section 3.1.

\section{Results and Analysis}

\subsection{0108+388}
 
The radio source 0108+388 is associated with a narrow emission 
line galaxy at z = 0.670 with R = 22.0 mag (Stickel et al. 1996a,
Stanghellini et al. 1993). This GHz-peaked radio source has been
classified as a compact symmetric object, 
with an inverted spectrum in the nucleus, plus twin-jets with steeper spectra
extending 3 mas from the nucleus toward the northeast and southwest (Taylor,
Readhead, and Pearson 1996, Conway et al. 1994).
Unlike most compact symmetric objects, this source also shows
large scale structure, having a 
steep-spectrum jet extending about 25$''$ to the east of the nucleus
(Baum et al. 1990). Optical images of the field show a very red,
diffuse, and `slightly asymmetric' galaxy, perhaps a face-on
spiral, although the faintness of the galaxy makes classification
difficult (Stanghellini et al. 1993).  
The radio source projects within 0.5$''$ of the center of this
galaxy. There is no evidence for a strong point source contribution
in the R band image of 0108+388.
The I band image is more compact, leading Stanghellini et al. (1993)
to propose an increased contribution from the active nucleus in the
near IR, perhaps indicating `the existence of nuclear 
obscuration' toward the AGN in 0108+388.
Monitoring by Stickel et al. (1996a) shows
variability of this source in the near IR, with a very 
red observed color during IR maximum ($\alpha_{IR}^{opt}$ $\le$ $-$3).

We have searched for HI 21cm absorption toward 0108+388 with the WSRT
centered at the heliocentric redshift of the parent galaxy.
The resulting spectrum  at the position of the peak radio continuum
surface brightness is shown in Figure 1. 
The upper frame shows the observed spectrum 
after subtracting the peak surface brightness of 0.146 Jy beam$^{-1}$, 
and after removing a linear baseline using a 
fit to off-line channels. The lower frame shows the spectrum 
converted to optical depth using the 
source continuum flux density. 

A strong HI 21cm absorption line is detected at z = 0.66847 with a
width of 100 km  s$^{-1}$ and a peak optical depth
of 0.44.  The continuum source is resolved with the
WSRT at 851 MHz,  with a
peak surface brightness of 146 mJy beam$^{-1}$ and a total flux density
of 180 mJy (Figure 2a). 
The contours in Figure 2a extend to the east, consistent with emission from
the large-scale jet seen by  Baum et al. (1990). 
The HI absorption line image shows no indication of extension of the 
absorption toward the east (Figure 2b), although the
sensitivity is such that we can only set a 1$\sigma$ limit  of 0.25 
to the optical depth toward the jet at a distance of 20$''$ east of
the nucleus ($\sigma \equiv$ spectral rms noise).

The absorption profile is fit reasonably well by a single Gaussian
component (the dashed-line in the lower frame of Figure 1),
whose parameters are listed in Table 2, although
we cannot preclude a profile composed of a few narrower, blended components.
Column 2 in Table 2  lists the heliocentric redshift corresponding to
zero velocity in the spectrum (the `reference redshift' $\equiv$
z$_{ref}$), and column 3 
lists the heliocentric redshift corresponding to 
the peak of the absorption component (z$_{abs}$). Column
4 lists the component velocity relative to the
reference redshift. Columns  5, 6, and 7 list the velocity integrated 
optical depth,
the peak optical depth, and the velocity full width at half maximum
(FWHM),  respectively. 
Column 8 lists the integrated HI column density for the component.
The integrated HI column density for this system
is 80$\times$10$^{18}$$\times({{T_s}\over{f}})$ cm$^{-2}$,
where T$_s$ is the spin
temperature of the gas and $f$ is the HI covering factor. 

Using the spectrum of the total radio emission from 0108+388,
and the spectrum of the core-jet emission on
milli-arcsecond scales (Baum et al. 1990), we
estimate that the milli-arcsecond scale structure in 0108+388
contributes about 70$\%$ to the
total flux density at 851 MHz, and about 85$\%$ to the peak
surface brightness as seen by the WSRT at 851 MHz.
Further, the compact nucleus has an inverted spectrum such that the
low frequency emission from the core-jet must be dominated by
the jet (Baum et al. 1990). 
Given the 44$\%$ optical depth observed in the HI 21cm line,
it appears likely that the HI  
covers most of the milli-arcsecond jet structure in 0108+388. The implied 
lower limit to the cloud size is then
about 6 mas, or 36 pc for H$_o$ = 65 km s$^{-1}$ Mpc$^{-1}$ and q$_o$
= 0.5. 

We have also searched for absorption by the 2cm
2$_{11}$-2$_{12}$ transition of H$_2$CO associated with the redshifted HI 21cm
absorption system toward 0108+388 using the Very Large Array operated by the
National Radio Astronomy Observatory on June 21, 1997. This line is
redshifted to 8683.692 MHz. The source continuum  flux density is 
0.943 Jy at this frequency. No absorption was detected 
to a 1$\sigma$ limit of 0.26 mJy per channel at a velocity resolution
of 6.8 km sec$^{-1}$, implying a 1$\sigma$ optical depth limit of 0.03$\%$.
This limit is comparable to the limit found for the same
molecular transition in the z = 0.6733 
~HI 21cm absorption line system seen toward 1504+377 (Carilli et
al. 1997).  

\subsection{0500+019}

The radio source 0500+019 shows an exponential cut-off in the spectrum going
from the near IR into the optical (Stickel et al. 1996b). The source is
also variable in the near IR, and in the radio (Stickel
et al. 1996a), and the radio source is 
unresolved at cm wavelengths at resolutions ranging from a few mas to
a few arcseconds (Taylor and Perley 1996, Hodges and Mutel 1984). 

A 2.2 $\mu$m image of this source reveals a point source
with faint emission extending 3$''$ to the south, with a
total magnitude of K = 15.9 mag. The R band image shows
an R = 20.7 mag inclined  disk or lenticular
galaxy with a major axis of about 4$''$ (Stickel et al. 1996b).  
This galaxy is associated with a
group of ten galaxies distributed over an area of about 60$''$.
There is no evidence for a point source in the R band image,
although the galaxy brightness profile is asymmetric, peaking about 
1$''$ to the north of center along the major axis of the galaxy, 
coincident within the errors with the position of the K band point source.

Optical spectra of this system reveal a moderate ionization, narrow
emission line spectrum (Stickel et al. 1996b). From four narrow emission
lines, and from the CaII H and K absorption lines, the redshift
for this galaxy is found to be z = 0.5834$\pm$0.0014, where the error 
indicates the full spread in estimated redshifts for all the lines. 
Stickel et al. (1996b) also detect a narrow emission line at 0.6354 $\mu$m
which cannot be reasonably identified with the z = 0.5834 galaxy.
Based on this unidentified emission line, on the off-center
location of the 2.2 $\mu$m point source relative to the galaxy center,
and on the asymmetric galaxy profile in the R band image, Stickel et
al. suggest that the quasar may be a background source at z $>$ 0.5834,
and that the red color of the quasar is due to dust in the foreground
galaxy. 

Our first WSRT spectrum of 0500+019 was
centered at the expected frequency for HI 21cm absorption at
the redshift of the galaxy, and an absorption feature was
observed at the edge of the bandpass at 896.4 MHz. 
We then re-observed this system,
centering the bandpass  on the HI 21cm absorption feature.
The resulting spectra are shown in Figure 3.
The upper frame shows the redshifted HI 21cm 
absorption spectrum for observations on
Nov. 12, 1996. The source continuum flux density of
1.6 Jy has been subtracted, and a linear baseline removed using
a fit to off-line channels. The middle frame shows
the spectrum of 0500+019 taken on Nov. 17, 1996.
(Note that in Figure 3 the channels for spectra from both days have been
aligned such that the zero point in velocity 
corresponds to a heliocentric redshift of 0.58457).
The bottom frame shows the summed spectrum from the two days
(weighted by integration time), and converted to optical depth using the 
source continuum flux density. The dashed-line in this frame is
a two component Gaussian model fit to the spectrum.

Broad HI 21cm absorption is observed toward 0500+019 
centered near z = 0.58472 and spanning 
about 140 km s$^{-1}$ in velocity with a peak optical depth of about
0.04. The observed spectrum is fit reasonably
well by two Gaussians, whose parameters are listed in Table 2.
The total column density (summed over both absorption components) is
6.2$\times$10$^{18}$$\times({{T_s}\over{f}})$ cm$^{-2}$.

\subsection{1504+377}

The flat spectrum radio source 1504+377 (K\"uhr et al. 1981) 
is located at the center of an inclined disk, or lenticular, 
galaxy with a moderate-excitation narrow emission line spectrum at z = 
0.674$\pm$0.001 (Stickel and K\"uhr 1994). 
The parent galaxy may be associated with a group of
galaxies, and has a close companion galaxy at a projected distance of 
4$''$ (Stickel and K\"uhr 1994).
The source shows evidence for flaring of its  near IR intensity,
although optical images of 1504+377  show no indication of a bright AGN 
(Stickel and K\"uhr 1994, Stickel et al. 1996a).

Recent radio spectroscopic observations of 1504+377 have revealed
strong HI 21cm absorption, and absorption by a number of molecular
species, at the redshift of the parent galaxy (Wiklind and
Combes 1996a, Carilli et al. 1997), suggesting that 
the very red color of this quasar is due to
extinction. In this case evidence favors extinction by 
gas in the quasar's host galaxy, and in particular by dust
in a circumnuclear torus associated with the AGN itself (Wiklind and
Combes 1996a, Carilli et al. 1997). The molecular absorption in this system
shows two distinct velocity components: a narrow line system (line
width $\le$ 10 km s$^{-1}$) at
z = 0.67150 and a broad line system (line width = 100 km s$^{-1}$) at
z = 0.6733. The velocity separation for the two 
systems is 330 km s$^{-1}$. The HI 21cm absorption is detected only
over the velocity range covered by the broad molecular absorption
line system, and the optical redshift of the galaxy corresponds to within
the errors with this broad line system.

The WSRT observations of 1504+377 had two primary goals. 
The first goal was to verify the lack of HI 21cm absorption at the redshift
of the narrow molecular absorption system. The single dish spectrum
taken with the NRAO 140 ft telescope was
corrupted by terrestrial interference very close in frequency  to the
narrow line system. The second goal 
was to obtain a more accurate line profile for 
the broad line system, since the single dish spectrum
had non-linear spectral baselines and a 
10$\%$ uncertainty in the line-to-continuum ratio due to difficulties
in determining the continuum source flux density. 

The WSRT spectrum of 1504+377 is shown in
Figure 4. The zero point of the velocity scale was chosen to
be consistent with Wiklind and Combes (1996a), corresponding to 
the heliocentric redshift of the narrow absorption line system at
z = 0.67150. 
The upper frame shows the observed spectrum 
after subtracting the peak surface brightness of 1.0 Jy beam$^{-1}$, 
and after removing a linear baseline using a 
fit to off-line channel. The lower frame shows the spectrum 
converted to optical depth using the 
source continuum flux density. 

The WSRT spectrum confirms the broad absorption line system
at z = 0.6733, and  the lack of HI 21cm absorption associated
with the narrow molecular line system at z = 0.67150. 
The rms noise in the WSRT spectrum is
14 mJy at a velocity resolution of 
6.9 km s$^{-1}$. Hence the 1$\sigma$ 
column density limit for HI absorption associated with the narrow molecular
line system is 0.2$\times$10$^{18}$$\times({{T_s}\over{f}})$
cm$^{-2}$. A detailed discussion of the possible implications of the 
paucity of neutral hydrogen associated with this molecular cloud at z
= 0.67150 can be found in Carilli et al. (1997).

Following Carilli et al. (1997), we have fit a three component Gaussian
model to the broad absorption line system toward  1504+377 (see Table 2).
The resulting velocities for the two high optical depth components
agree to within 1 km s$^{-1}$ with results from the 140 ft telescope.
However, the opacities are about 20$\%$ higher in the WSRT spectrum. 
This latter difference can be attributed to the more accurate, 
and somewhat lower, flux density for the background source,
and a more linear spectral baseline in the WSRT spectrum.
The integrated HI column density for the system (summed over the three
absorption components) is 
42$\times$10$^{18}$$\times({{T_s}\over{f}})$ cm$^{-2}$.

\subsection{2149+056}

The 2149+056 radio source is associated with an R = 20.4 mag
emission line galaxy at
z = 0.740 (Stickel and Kuhr 1993). Stickel et al. (1996a)
find evidence for a very steep optical spectrum and moderate variability
in the near IR. They suggest that 
this object is an `IR variable quasar within a luminous galaxy.'
The radio source is unresolved at cm wavelengths at resolutions ranging from
0.15$''$ to a few arc seconds (Taylor and Perley 1996).

We searched for HI 21cm absorption at z = 0.740 toward 2149+056
using the WSRT. The resulting spectrum is shown in Figure 5, 
after subtracting the peak surface brightness of 0.50 Jy beam$^{-1}$, 
and after removing a linear baseline using a 
fit to all the channels.

No absorption is seen to an optical depth limit (1$\sigma$)
of  0.03 at 7.2 km s$^{-1}$ resolution. We have also smoothed the
spectrum to lower velocity resolution and find no absorption 
to an optical depth limit of 0.015 at 29 km s$^{-1}$ resolution.
The corresponding column density limit is
N(HI) $\le$ 0.4$\times$10$^{18}$$\times({{T_s}\over{f}})$ cm$^{-2}$
for a narrow line (7 km s$^{-1}$), or 
0.8$\times$10$^{18}$$\times({{T_s}\over{f}})$ cm$^{-2}$
for a broad line  (29 km s$^{-1}$).

\section{Discussion}
\subsection{Reddening toward Red Quasars}

We have searched for redshifted neutral hydrogen 21cm absorption toward
five sources from the Stickel et al. (1996a) red quasar sub-sample
with the new UHF system at the WSRT. 
Three of the five show 21cm absorption (0108+388, 0500+019,
1504+377), while no absorption was seen 
toward 2149+056 at a level below that seen for the three 
detections. The WSRT spectra of 0202+149 were corrupted by terrestrial
interference. A sixth source on the Stickel et al. red quasar
list, 0218+357, had been
detected previously in HI 21cm absorption at z = 0.6847 (Carilli, Rupen,
and Yanny 1993), and we include this source in the discussion below.

The average value of the radio-to-optical spectral index, 
$\alpha_{rad}^{opt}$, for optically identified, flat
spectrum radio loud quasars is:~  $\alpha_{rad}^{opt}$ = $-0.6$
(Condon et al. 1983), while the values for the red quasar sub-sample are:
$\alpha_{rad}^{opt}$ $<$ $-0.9$. 
On the other hand, the radio-to-near IR spectral indices 
for most of the red quasar sub-sample are: $\alpha_{rad}^{IR}$
$\ge$  $-0.8$, implying signficant steepening of the spectra
between the near IR and the optical:~
$\alpha_{IR}^{opt}$ $<<$ $-1$ (Stickel et al. 1996a). 
A rough lower limit to the required extinction can be derived by
comparing the observed optical flux density with that predicted by
extrapolating the radio-to-IR spectrum into the optical. This is a
lower limit due to confusion by emission from
stars in the galaxy with which the absorbing gas is 
associated. Values of the rest frame visual
extinction, A$_V$, range from: ~A$_V$ $\ge$ 2 for 
0108+388 and 0500+019, to A$_V$ $\ge$ 3 for 0218+357 (Menten and
Reid 1996), to  A$_V$ $\ge$ 7 for 1504+377. 
Using the observed value of N(HI) derived from the HI 21cm absorption lines
leads to  lower limits to the spin temperatures of about 50 K for
0108+388, 300 K for 1504+377, 500 K for 0500+019, and 1000 K for 0218+357,
assuming a Galactic dust-to-gas ratio (Spitzer 1978), and $f~=~1$. 

The large neutral hydrogen columns  seen toward these four red quasars
provides circumstantial evidence in favor of the reddening hypothesis,
as opposed to an intrinsically red spectrum for the AGN emission
mechanism. While we cannot rule-out the latter hypothesis based
on the HI absorption data alone,
the required dust-to-gas ratios would have to be less than
about  0.1 to 0.3 $\times$ Galactic in order to reduce the A$_V$
values significantly below unity for reasonable values of T$_s$. 

Of the five red quasars from the Stickel et al. (1996a) sample
that have been searched to reasonable optical
depths in HI 21cm absorption, the only source that is
not detected in absorption is 2149+056, leaving open the
possibility of an intrinsically red emission spectrum for this source.
However, Stickel et al. (1996a) point out that the
apparent magnitude of the
`parent' galaxy for 2149+056 is anomalously bright given the redshift
of 0.740, and they raise the  possibility that
the emission line spectrum is that of the quasar, while the
galaxy, and hence the reddening material, 
may be at a redshift less than 0.740. Deep optical spectroscopy of the
parent galaxy could  test this possibility.

We have detected high column density
absorption systems in four of five sources from the radio 
selected red quasar sub-sample. This detection rate is significantly
higher than in optically selected samples. For instance, 
two independent searches for HI 21cm absorption associated with optically
selected MgII absorption line systems resulted in two detections out
of 18 systems in both studies (Briggs and Wolfe 1983, Lane et
al. 1997), where the optical depth limits in these studies 
were comparable to the limits presented herein.

\subsection{The Statistics of Red Quasars}

We now consider the question of biases in
optically selected samples of quasars due to dust obscuration, 
using the  statistics for the  1 Jy sample of flat spectrum radio
sources, and the high detection rate of HI 21cm absorption toward 
red quasars. In the following section, the term `optically
selected sample' means a sample of quasars selected from a 
moderately deep optical survey such as the POSS, with a redshift
distribution comparable to the 1 Jy quasar sample. 

The Stickel et al. (1996a) red quasar sub-sample
was drawn from 299 flat spectrum sources in the complete, all-sky
1 Jy sample of radio sources (K\"urh et al. 1981), 
with the additional constraint of
declination greater than -20$^o$. Correcting for the area of the sky
covered then implies a total of about 200 sources, of which Stickel et
al. find that 81.5$\%$ are quasars and
11$\%$ are galaxies. The remaining 7.5$\%$ are unidentified on the
POSS (the 15 `red quasars'), implying 
a magnitude limit of:~ $m$ $\ge$ 20, or~ $\alpha_{rad}^{opt}$ $<$ $-0.9$.
Assuming that the 80$\%$ detection rate of HI 21cm absorption toward
the five sources discussed herein applies to the red 
quasar sub-sample as a whole, leads to a lower limit 
to the fraction of quasars
`missing' from an optically selected sample 
due to dust obscuration of: ~ ${0.8\times5}\over{200}$ = 6$\%$.
An upper limit to this fraction can be estimated by making the
extreme assumption that all the red quasars and the galaxies in the
sample are dust-obscured quasars. This leads to an upper limit 
to the fraction of missing quasars of:~ ${22+15}\over{200}$ = 20$\%$.

Fall and Pei (1993) have 
presented detailed models of the numbers of quasars missing from 
optically selected samples due to obscuration by dust associated with
high column density quasar absorption line systems as a function of
redshift. The redshift distribution of the 1 Jy, flat spectrum quasar
sample shows a broad peak at z $\approx$ 1, with a tail to higher
redshifts (Stickel, Meisenheimer, and K\"uhr 1994). 
Using this redshift distribution, the models of Fall and Pei 
predict that between 2$\%$ and 12$\%$ of the sources would be missing
from such a sample due to dust obscuration. 


We can also make a rough estimate of how many high column density
quasar absorption line systems  are
missing from optically selected quasar samples. The statistics of damped
Ly$\alpha$ systems in optically selected samples implies a redshift
distribution function of the form: N(z) = 0.015 $\times$ (1+z)$^{2.3}$, where
N(z) is the number of absorption systems with 
N(HI) $\ge$ 2$\times$10$^{20}$ cm$^{-2}$ per unit redshift
(Rao, Turnshek, and Briggs 1995). Combining 
this function with the redshift distribution of the Stickel
etal. quasar sample leads to an expected number of 17 high column density
absorption systems  toward their optically identified quasars
(ie. out of about 163 sources). 
Again, assuming that the 80$\%$ detection rate of HI 21cm absorption
toward the five red quasars studied herein applies to the red 
quasar sub-sample as a whole, leads to a lower limit to
the fraction of high column density absorption systems missing from 
optically selected samples of:~
${0.8\times15}\over{17+0.8\times15}$ = 40$\%$.
The upper limit to this fraction is:~ ${15+22}\over{15+22+17}$
= 70$\%$, again determined by making the extreme assumption that 
all the galaxies and red quasars in the Stickel et al. flat spectrum
source sample have high
column density absorption systems. The comparative numbers 
from the models of  Fall and Pei (1993) are between 30$\%$ and
60$\%$ (see also Fall and Pei 1995).

We should emphasize that there are a number of significant 
uncertainties in the statistics presented above. 
First is simply the small number of red quasars searched thus far for
redshifted HI 21cm absorption. 
Second are the possible biases introduced
by the fact that, due to
practical observational limitations, our absorption searches were
limited to the specific redshifts of the parent galaxies of the 
absorbing clouds, as determined from emission lines seen
in  deep optical spectra (Stickel etal. 1996a). 
It is possible that we have under-estimated the number
of high column density systems by not searching over the full
redshift range in each case ({\sl eg.} absorption systems associated with 
dwarf galaxies might be missed).
Conversely, this selection criterion might skew the results toward
absorbers at the redshift of the quasar host galaxy 
(see section 4.3), which perhaps could bias the statistics in the
opposite sense. 
And third is the fact that the fairly high radio flux density 
limit of the 1 Jy sample
biases the sample toward lower redshift quasars: the redshift
distribution of the 1 Jy quasar
sample peaks at z $\approx$ 1, while quasar samples derived from
radio surveys with 
lower flux density limits peak at z $\approx$ 2, close to the 
redshift peak found for optically selected quasar samples (Shaver et
al. 1996, Hewitt and Burbidge 1987). 
While the redshift distribution is used explicitly
in the calculations above, unbiased 
searches for redshifted HI 21cm absorption
toward a sample of red quasars selected from fainter radio catalogs
would make for a fairer comparison with the optical data. 


Overall, the statistics presented above should not be considered
rigorous, but only representative. Still, the red quasar data
support the basic conclusions of Fall and Pei (1993):
(i) there may be a significant, but not dominant, population of 
quasars missing from optically selected samples due to
obscuration, perhaps as high as 20$\%$ at the POSS limit
for an optical sample 
with a redshift distribution similar to the 1 Jy, flat spectrum quasar
sample, and (ii) optically
selected samples may miss about half the high column density
quasar absorption line systems.
A final point made by Fall and Pei is that the strongest bias in  optical
samples  may be against the high column density systems with
high metalicities and  dust-to-gas ratios. 
Such systems might be the most interesting 
from the perspective of follow-up radio studies 
of the dense, pre-star forming ISM in galaxies at significant look-back
times. The data presented herein suggests that 
searching for redshifted HI 21cm absorption toward red quasars may be
an effective method for circumventing this bias.


\subsection{The Location of the Absorbing Material}

In some sources in our study the absorption, and presumably the
reddening, is due to material in the quasar's host
galaxy, as appears to be the case 
for 1504+377 at z = 0.673 (Carilli et al. 1997, Wiklind and Combes
1996a), and for 0108+388 at z = 0.6685.
In 1504+377 and 0108+377 the velocity
widths for the absorption systems are large ($\approx$100 km
s$^{-1}$), and hence comparable to line widths
seen for systems associated with nearby Seyfert
galaxies and other low redshift AGN (Dickey 1983, 
van Gorkom et al. 1989, Conway
and Blanco 1995). High resolution imaging of the
absorption toward a number of nearby AGN indicates that 
the high velocity dispersion gas is local to the AGN, perhaps
in a circumnuclear torus or ring with size scale of order 10 pc,  and
that the velocity width reflects the torus dynamics (Israel et
al. 1991, Gallimore et al. 1994, Mundell et al. 1995a, Mundell et
al. 1995b, Conway 1997).  

In other red quasars  the absorption is
due to material in galaxies cosmologically distributed along the line of
sight to the quasar. One such intervening system is the 
gravitational lens 0218+357, the smallest ``Einstein ring'' radio
source, for which HI 21cm and molecular absorption has been detected
toward the radio source at z = 0.687
(Carilli, et al. 1993, Wiklind and Combes 1995,
Menten and Reid 1996).  In retrospect, gravitational lensing may not
be surprising for such systems, since one requires a low impact-parameter
line of sight  in order to intercept the denser regions of the ISM of the
intervening galaxy. An important conclusion is that 
at least some gravitational lenses are  gas rich systems, 
as has also been found in the case of the Einstein ring radio source
PKS 1830-211 (Wiklind and Combes 1996b, Frye, Welch, and Broadhurst
1997, Menten, Carilli, and Reid 1997).

The case for a cosmologically 
intervening absorption system versus an associated
system for  0500+019 is unclear (see section 3.2). 
A possible argument against the intervening 
hypothesis for 0500+019 is the broad HI 21cm absorption line profile
(total velocity width = 140 km s$^{-1}$). 
This is much larger than  the 10 km s$^{-1}$ 
to 20 km s$^{-1}$ velocity 
dispersion expected for a line-of-sight perpendicular to the 
disk of a spiral galaxy (Dickey and Lockman 1989), and might argue
for absorption by circumnuclear gas, as discussed above. 
An additional puzzle for this system is that 
the HI redshift is offset from the optical redshift by  +240$\pm$265 
km s$^{-1}$ (i.e., in-falling relative to the optical galaxy),
although the uncertainty in the optical redshift is large.
If the absorber is an intervening system, 
a possible model to explain both the large velocity offset and the large
velocity dispersion  may be galactic rotation, in the case where the
quasar projects off-center along the major axis of a highly inclined 
disk galaxy. For example, if the galaxy 
rotation velocity was 200 km sec$^{-1}$,
and the line-of-sight to the quasar cut through the edge-on disk halfway out
along the major axis, the expected line width due to differential
rotation would be about 100 km s$^{-1}$. One possible 
difficulty with this model is that HI 21cm absorption
seen through the disk of the Milky Way galaxy typically appears as
a series of discrete, narrow components with widths of a few km
s$^{-1}$ (Dickey et al. 1983).
Unfortunately, the sensitivity of the current WSRT spectrum of
0500+019 is insufficient to differentiate between a blend of narrow
components, or two broad, shallow lines.

\vskip 0.2truein 

We thank A.G. de Bruyn, Tony Foley, A.-J. Boonstra, and the staff
at the WSRT for their assistance with these observations, 
and M. Fall, P. Shaver, G. Taylor, F. Briggs,
and the referee, J. van Gorkom,  for useful comments and discussions.
This research made use of the NASA/IPAC Extragalactic Data Base (NED)
which is operated by the Jet propulsion Lab, Caltech, under contract
with NASA.

\newpage

\centerline{\bf References}

Baars, J., Genzel, R., Pauliny-Toth, I., and Witzel, A. 1977, A.\&A.,
61, 99

Baum, S.A., O'Dea, C.P., Murphy, D.W., and de Bruyn, A.G. 1990, A\&A,
232, 19



Bregman, J.N.,Lebofsky, M.J., Aller, M., Reike, G.H., Aller, H.,
Hodge, P., Glassgold, A., and Huggins, P. 1981, Nature, 293, 714

Briggs, F.H. and Wolfe, A.M. 1983, Ap.J., 268, 76
 




Boonstra, A.-J. 1996, NFRA Newsletter, 11, 4


Carilli, C.L., Menten, K.M., Reid, M.J., and Rupen, M.P. 1997, 
Ap.J. (letters), 474, L89

Carilli, C.L., Lane, Wendy, de Bruyn, A.G., Braun, R., and Miley,
G.K. 1996, A.J., 111, 1830


Carilli, C.L., Perlman, E.S., and Stocke, J.T. 1992, Ap.J. (letters),
400, L13

Carilli, C.L., Rupen, M.J., and Yanny, Brian 1993, Ap.J., 412, L59

Clark, B.G. 1980, A.\&A., 89, 377

Condon, J.J., Condon, M.A., Broderick, J.J., and Davis, M.M. 1983,
A.J. 88, 20




Conway, J.E. and Blanco, P.R. 1995, Ap.J. (letters), 499, 131

Conway, J.E., Myers, S.T., Pearson, T.J., Readhead, A.C., Unwin, S.C.,
and Xu, W. 1994, Ap.J. 425, 568

Conway, J.E. 1997, in {IAU Colloquium 164: Radio Emission from
Galactic and Extragalactic Compact Radio Sources}, eds. J.A. Zensus,
J. Wrobel, and G.B. Taylor (Berkeley, PASP), in press.

Dickey, J.M. and Lockman, F.J. 1990, A.R.A.A., 28, 215

Dickey, J.M. 1983, Ap.J., 263, 87

Dickey, J.M., Kulkarni, S.R., Heiles, C.E., and van Gorkom, J.H. 1983,
Ap.J. (Supplement), 53, 591


Fall, S.Michael and Pei, Yichuan C. 1995, in {\sl QSO Absorption
Lines}, ed. G.Meylan, (Springer: Heidelberg), p. 23

Fall, S.Michael and Pei, Yichuan C. 1993, Ap.J., 402, 479

Frye, Brenda, Welch, William J., and Broadhurst, Tom 1997,
Ap.J. (letters), 478, L25

Gallimore, J.F., Baum, S.A., Odea, C.P., Brinks, E., and Pedlar,
A. 1994, Ap.J. (letters), 422, L13

Heisler, J. and Ostriker, J. 1988, Ap.J., 332, 543

Hewitt, A.  and Burbidge, G. 1987, Ap.J. (Supplement), 63, 1


Hodges, M.W. and Mutel, R.L. 1984, A.J., 89, 1391

Israel, F.P., van Dishoeck, E.F., Baas, F., Koornneef, J., Black, J.,
and de Graauw. T. 1990, A\&A, 227, 342

K\"uhr, H., Witzel, A., Pauliny-Toth, I.I., and Nauber, U. 1981, 
A\&A (Supplement), 45, 367

Lane, W., Briggs, F.H., and Smette, A. 1997, in {\sl Probing the
Intergalactic Medium with Quasar Absorption Lines,} ed. P. Petitjean,
in press.



 

 
Menten, K.M., Carilli, C.L., and Reid, M.J. 1997, Ap.J. (letters),
in preparation

Menten, K.M. and Reid, M.J. 1996, Ap.J. (letters), 465, L99


Mundell, C.G., Pedlar, A., Edlar, A., Baum, S.A., Odea, C.P.,
Gallimore, J.F., and Brinks, E. 1995a, M.N.R.A.S., 272, 355

Mundell, C.G., Pedlar, A., Axon, D.J., Meaburn, J., and Unger,
S.W. 1995b, M.N.R.A.S., 277, 641






Rao, S., Turnshek, D., and Briggs, F.H. 1995, Ap.J., 449, 488

Shaver, P.A., Wall, J.V., Kellermann, K.I., Jackson, C.A., and
Hawkins, M.R.S. 1996, Nature,  384, 439

Spitzer, Lyman Jr. 1978, {\sl Physical Processes in the Interstellar Medium},
(Wiley: New York)

Stanghellini, C., O'dea, C.P., Baum, S.A., and 
Laurikainen, E. 1993, Ap.J. (Supplement), 88, 1

Stickel, M., Meisenheimer, K., and K\"uhr, H. 1994, A\&A (Supplement),
105, 211

Stickel, M. and K\"uhr, H. 1994, A\&A (Supplement), 105, 67

Stickel, M., Rieke, G.H., K\"uhr, H., and Rieke, M.J., 1996a, 
Ap.J., 468, 556

Stickel, M., Rieke, M.J., Rieke, G.H, and K\"uhr, H. 1996b, 
A\&A, 306, 49


Taylor, G.B. and Perley, R.A. 1996, VLA Calibration Manuel, (NRAO:
Socorro)

Taylor, G.B., Readhead, A.C., and Pearson, T.J. 1996, Ap.J., 463, 95

van Gorkom, J.H., Knapp, J.H., Ekers, R.D., Ekers, D.D., Laing, R.A.,
and Polk, K.S. 1989, A.J., 97, 708

Webster, R.L., Francis, P.J., Peterson, B.A., Drinkwater, M.J., and
Masci, F.J. 1995, Nature, 375, 469

Wiklind, Tommy and Combes, Francoise 1996a, A\&A, 315, 86

Wiklind, Tommy and Combes, Franciose 1996b, Nature, 379, 139

Wiklind, Tommy and Combes, Franciose 1995, A\&A, 299, 382











\newpage

\centerline{Figure Captions}

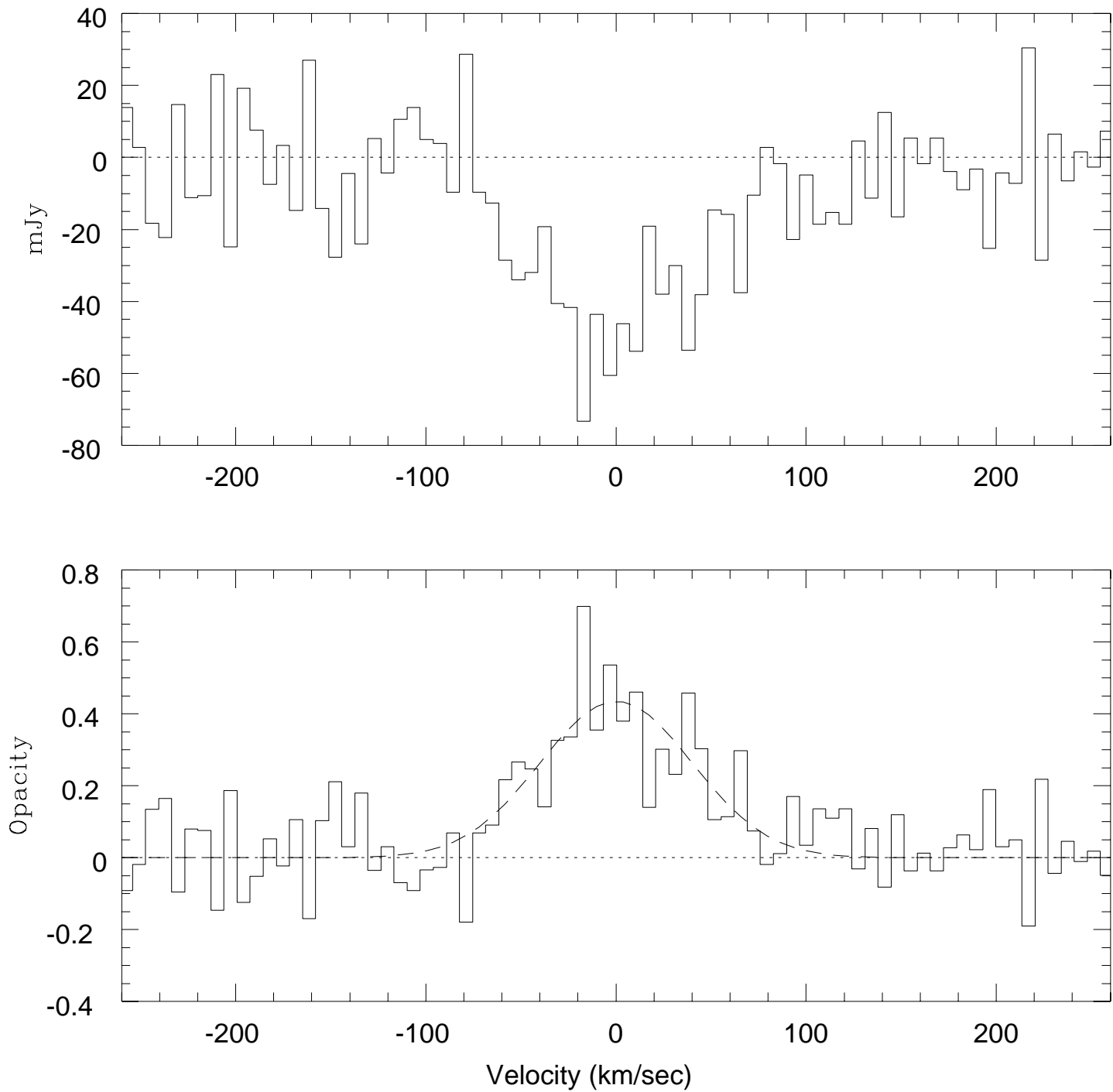
\noindent Figure 1 -- The upper frame shows the redshifted HI 21cm  
absorption spectrum toward 0108+388 for WSRT  observations on
Jan. 10, 1996. The Gaussian restoring beam for this observation was
FWHM = 38$''$x23$''$, with the major
axis oriented north-south. The zero point of the velocity
scale corresponds to a heliocentric redshift of 0.66847 for the HI
21cm line. The peak surface brightness of 0.146 Jy beam$^{-1}$ 
has been subtracted, and a linear baseline removed using
a fit to off-line channels.
The bottom frame shows the HI optical depth spectrum plus 
a single component Gaussian model fit (dashed-line). The parameters
for the fit in this and subsequent figures are given in Table 2. 

\noindent Figure 2 -- Figure 2a (upper) shows a radio continuum image
of 0108+388 at 851 MHz with a Gaussian restoring beam of 
FWHM = 38$''$x23$''$, with the major
axis oriented north-south. The contours levels are:
-20, -14, -10, 10, 14, 20, 28, 40, 57, 80, and 113 mJy beam$^{-1}$
and the peak surface brightness is 146 mJy beam$^{-1}$.
Figure 2b (lower) shows a continuum subtracted
spectral channel image for redshifted HI 21cm absorption 
towards 0108+388 at z$_\odot$ = 0.66847, corresponding to the
peak optical depth in HI with a velocity resolution of 6.9 km
s$^{-1}$. The contour levels are: $-$64, $-$48, $-$32, $-$16, 
16, 32, 48, and 64 mJy beam$^{-1}$. Dashed contours are negative. 
The cross indicates the position of the peak  
surface brightness in the continuum image in Figure 2a.

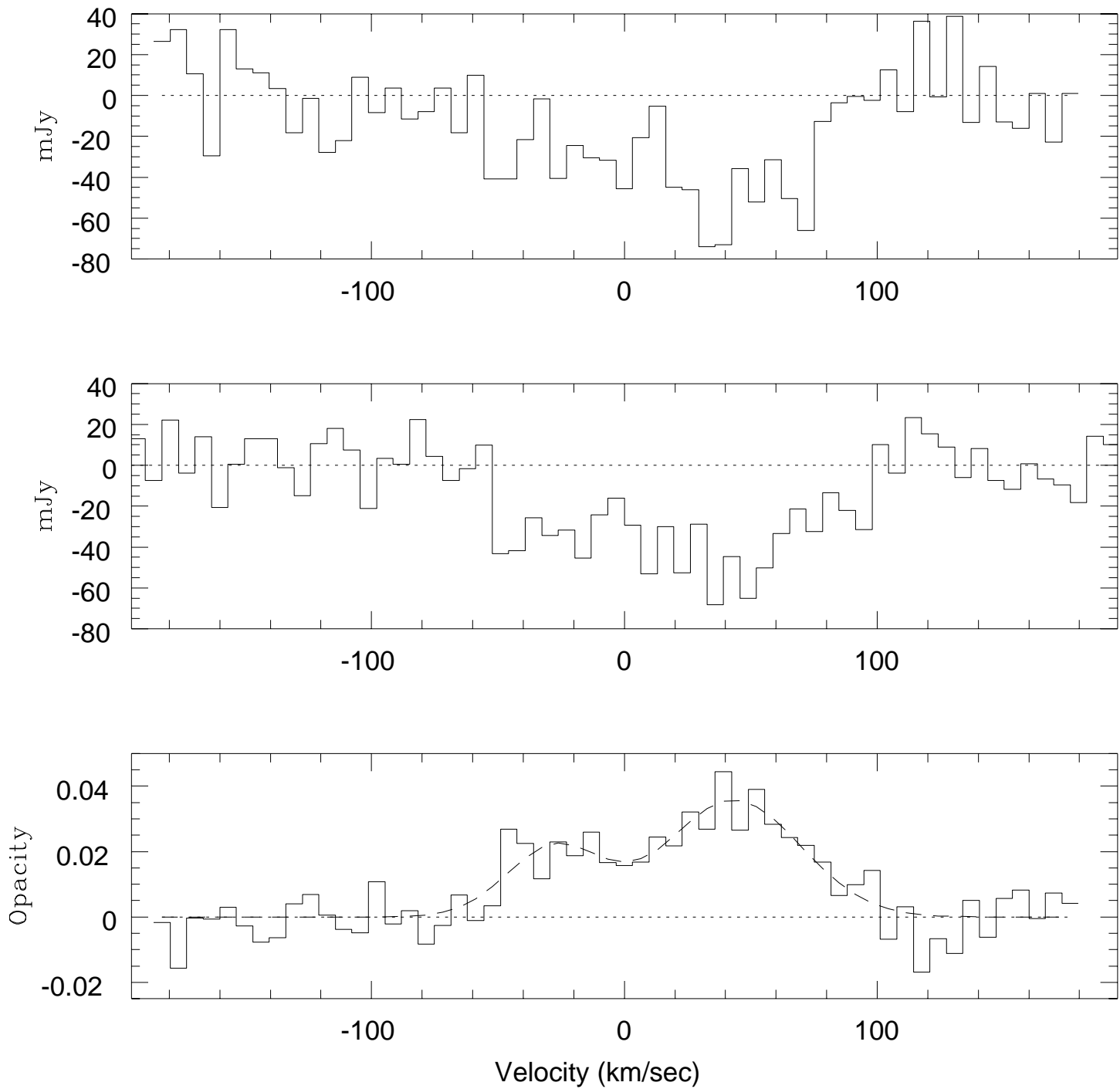
\noindent Figure 3 -- The upper frame shows the redshifted HI 21cm  
absorption spectrum of 0500+019 for WSRT observations on
Nov. 12, 1996. The Gaussian restoring beam for this observation was
FWHM = 660$''$x23$''$, with the major
axis oriented north-south. The zero point of the velocity
scale corresponds to a heliocentric redshift of 0.58457 for the HI
21cm line. The source continuum flux density of
1.6 Jy has been subtracted, and a linear baseline removed using
a fit to off-line channels. The middle frame shows
a similar spectrum, only taken on Nov. 17, 1996.
The bottom frame shows the summed spectrum from the two days
(weighting by integration time), and converted to optical depth
using the continuum flux density of the source, plus 
a two component Gaussian model fit (dashed-line).

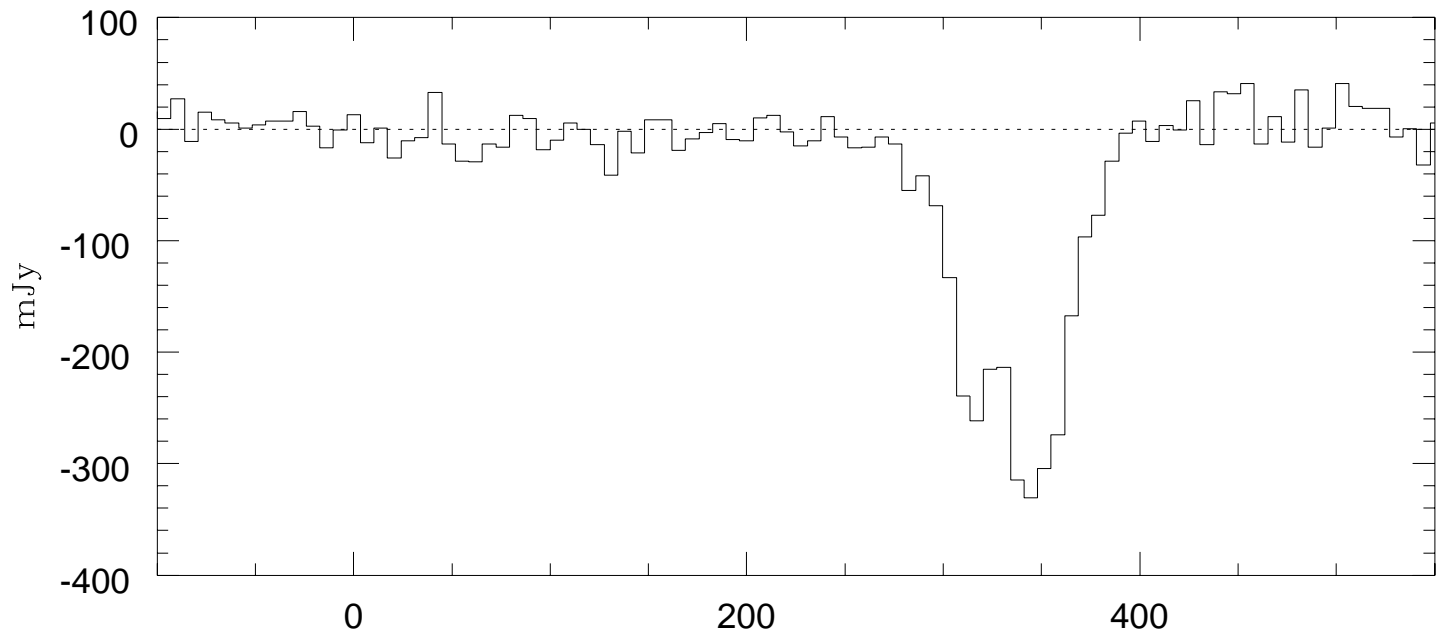
\noindent Figure 4 -- The upper frame shows the redshifted HI 21cm  
absorption spectrum of 1504+377 for WSRT  observations on
Nov. 10, 1996. The Gaussian restoring beam for this observation was
FWHM = 39$''$x23$''$, with the major
axis oriented north-south. The zero point of the velocity
scale corresponds to a heliocentric redshift of 0.67150 for the HI
21cm line. The source continuum flux density of
1.0 Jy has been subtracted, and a linear baseline removed using
a fit to off-line channels.
The bottom frame shows the HI optical depth spectrum plus 
a three component Gaussian model fit (dashed-line).

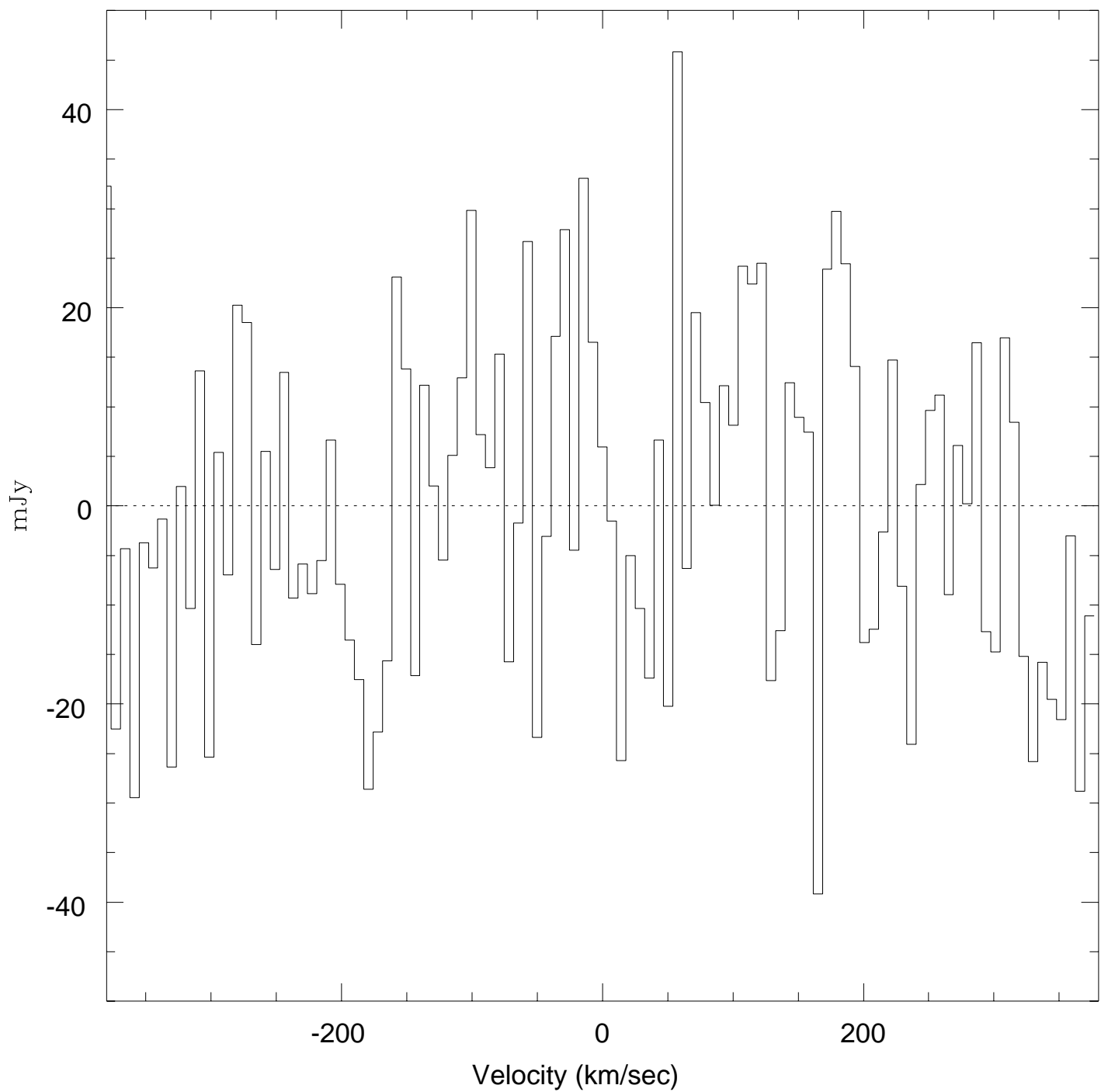
\noindent Figure 5 -- The redshifted HI 21cm  
absorption spectrum toward 2149+056 for WSRT  observations on
December 29, 1996. The Gaussian restoring beam for this observation was
FWHM = 220$''$x23$''$, with the major
axis oriented north-south. The zero point of the velocity
scale corresponds to a heliocentric redshift of 0.74000 for the HI
21cm line. The continuum flux density of 0.50 Jy
has been subtracted, and a linear baseline removed using 
a fit to all channels.

\newpage

\begin{deluxetable}{cccrccccc}
\footnotesize
\tablecaption{Observational Parameters \label{tbl-1}}
\tablewidth{0pt}
\tablehead{
\colhead{Source} & \colhead{Date}  & 
\colhead{Frequency} & \colhead{S$_\nu$} 
& \colhead{t$_{int}$} &  \colhead{Bandwidth}  &
\colhead{channels} & \colhead{Resolution}  & \colhead{rms}  \nl
\colhead{B1950} & \colhead{~}   & \colhead{MHz} 
& \colhead{Jy} & \colhead{hours}  & \colhead{MHz} & \colhead{~} &
\colhead{km s$^{-1}$} & \colhead{mJy beam$^{-1}$} 
}
\startdata
0108+388 & Jan 10 1997 & 851.037 & 0.18$\pm$0.02 & 
12 & 2.5 & 127 & 6.9 & 15 \nl
0202+149 & Nov~ 7 1996 & 774.910 & -- & 7 & 2.5 & 127 & 7.6 & -- \nl
0500+019 & Nov 12 1996 & 896.398 & 1.6$\pm$0.16 & 3 & 1.25 & 63 & 6.5
& 20 \nl
0500+019 & Nov 27 1996 & 896.350 & -- & 8 & 2.5 & 127 & 6.5
& 11 \nl
1504+377 & Nov 10 1996 & 849.017 & 1.0$\pm$0.1 & 12 & 2.5 & 127 & 6.9
& 14 \nl
2149+056 & Dec 29 1996 & 816.320 & 0.50$\pm$0.05 & 
11 & 2.5 & 127 & 7.2 & 16 \nl
\enddata
\end{deluxetable}

\newpage

\begin{deluxetable}{ccrrcrrc}
\footnotesize
\tablecaption{Gaussian Model Parameters \label{tbl-1}}
\tablewidth{0pt}
\tablehead{
\colhead{Source} & \colhead{z$_{ref}$} & \colhead{z$_{abs}$} & 
\colhead{Velocity} & \colhead{Int. Opt. Depth} & \colhead{Opt. Depth} & 
\colhead{FWHM}  & \colhead{N(HI)} \nl
\colhead{~} & \colhead{~} & \colhead{~} & \colhead{~~km s$^{-1}$}   
& \colhead{~km s$^{-1}$} 
& \colhead{~} & \colhead{~~km s$^{-1}$}  
& \colhead{~~$\times$${T_s}\over{f}$10$^{18}$cm$^{-2}$}
}
\startdata
0108+388 & 0.66847 & 0.66847$\pm$0.00003 & 0$\pm$4 & 46$\pm$7 & 0.44$\pm$0.04
& 94$\pm$10 & 80.7 \nl
0500+019 & 0.58457 & 0.58480$\pm$0.00002 & 43$\pm$3 & 2.5$\pm$0.3 & 0.036$\pm$0.003
& 62$\pm$7 & 4.5 \nl
0500+019 & -- & 0.58442$\pm$0.00003 & $-$27$\pm$5 & 1.4$\pm$0.3 
& 0.027$\pm$0.003 & 45$\pm$9  & 2.5 \nl
1504+377 & 0.67150 & 0.67331$\pm$0.00007 & 325$\pm$19 & 7.0$\pm$4.3 & 
0.073$\pm$0.045 & 85$\pm$22 & 12.8  \nl
1504+377 & -- & 0.67324$\pm$0.00001 & 313$\pm$1 & 4.0$\pm$0.6 & 0.22$\pm$0.03
& 16$\pm$2  & 7.3 \nl
1504+377 & -- & 0.67343$\pm$0.00001 & 347$\pm$1 & 13.4$\pm$4 & 0.34$\pm$0.08
& 34$\pm$5 & 24.4 \nl 
2149+056 & 0.74000 & 0.740$\pm$0.0023 & -- & $<$0.5 & $<$0.015 & 29 & $<$0.8 \nl
\enddata
\end{deluxetable}








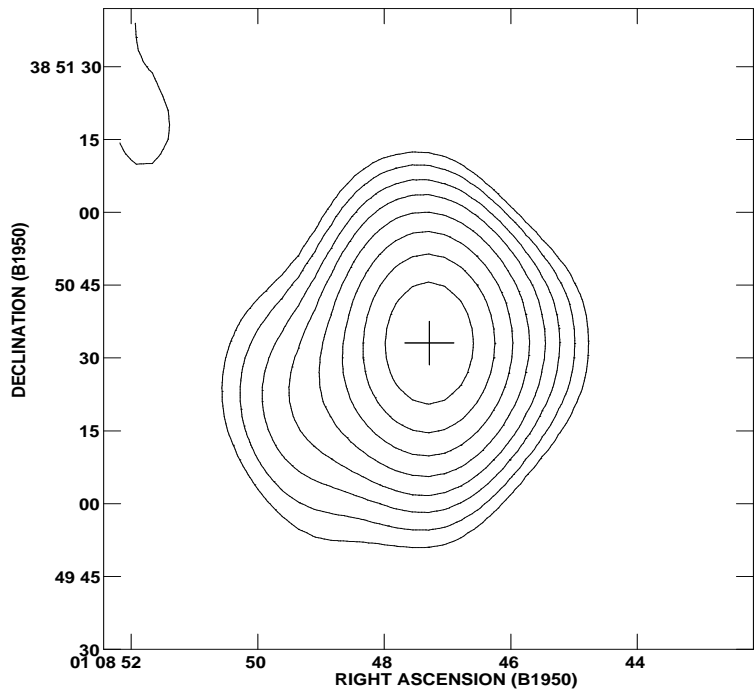
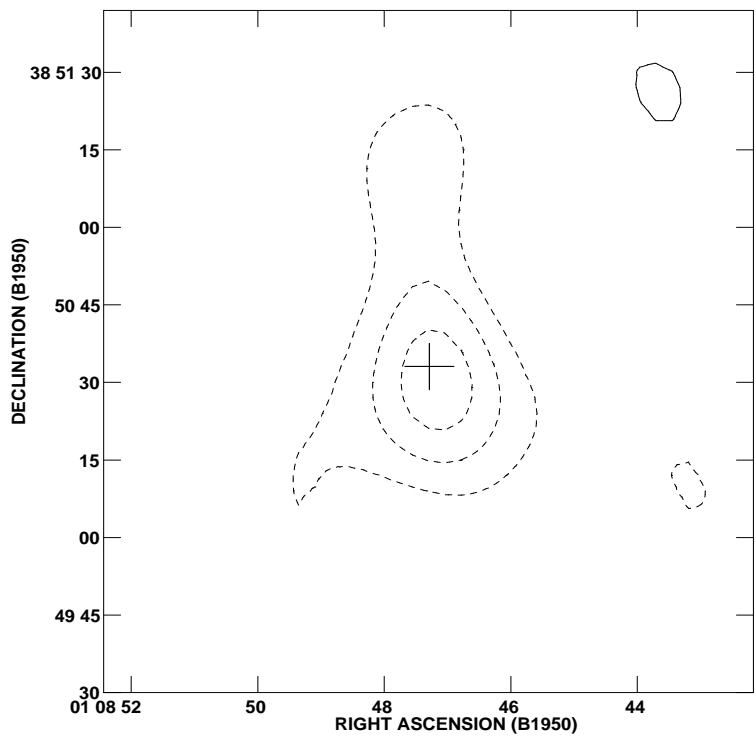
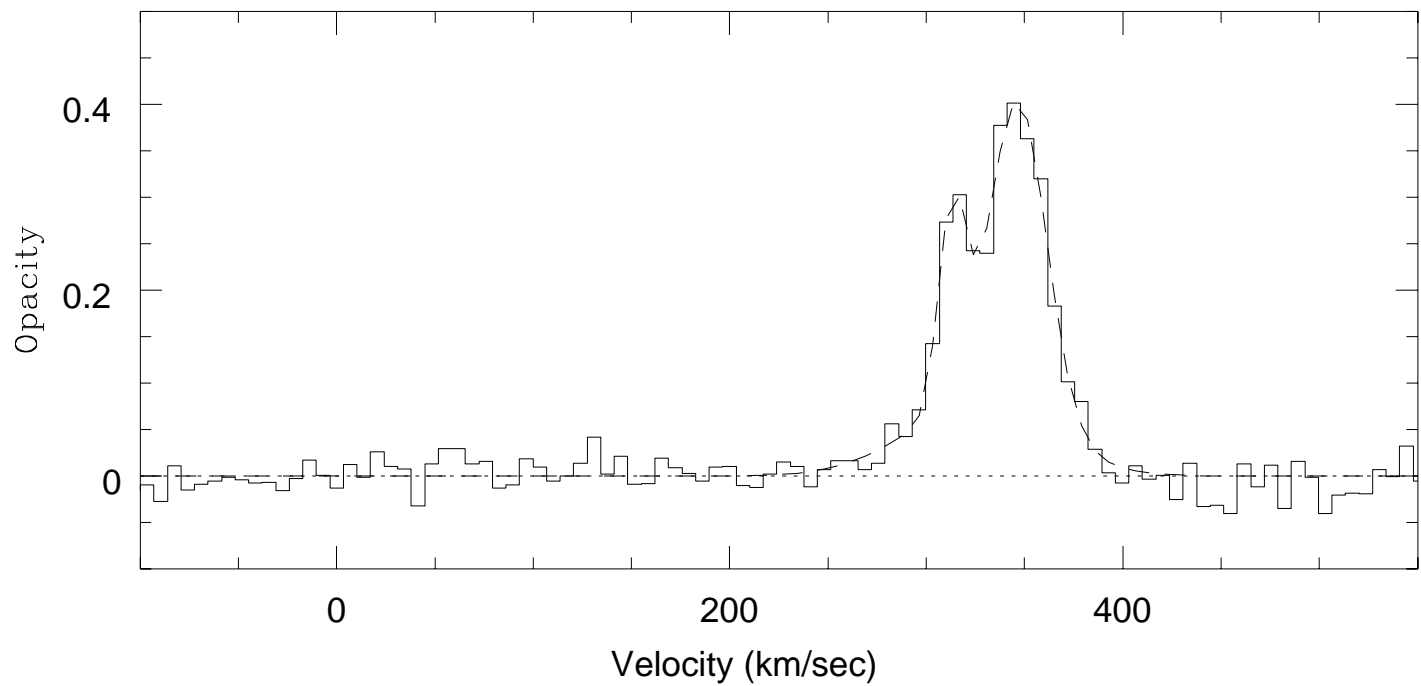
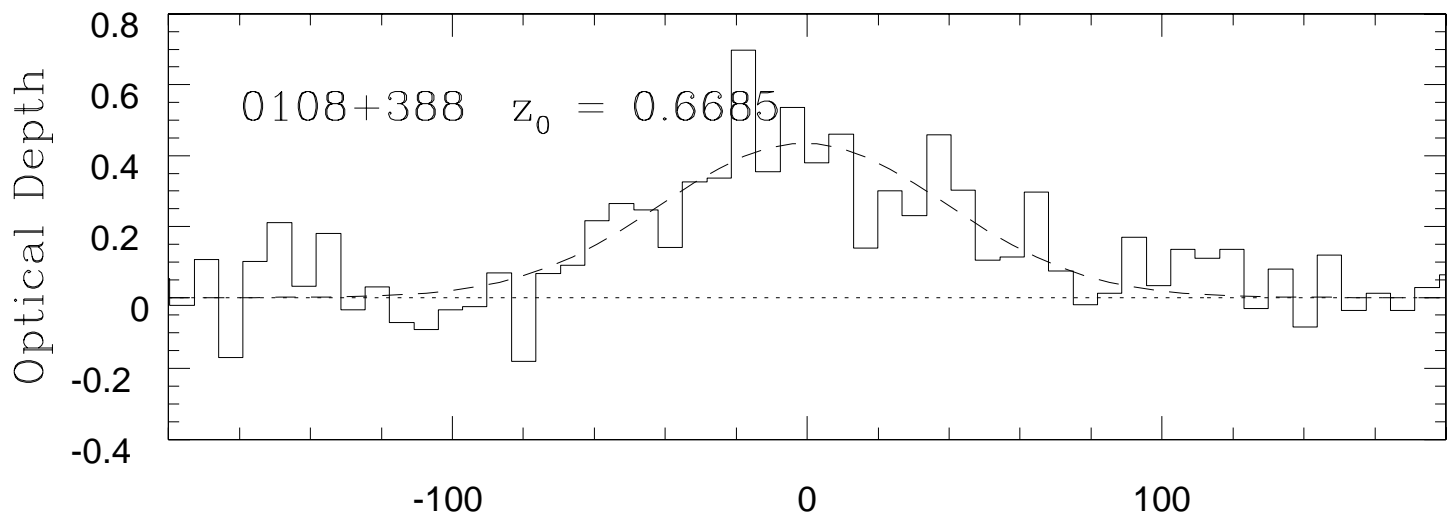
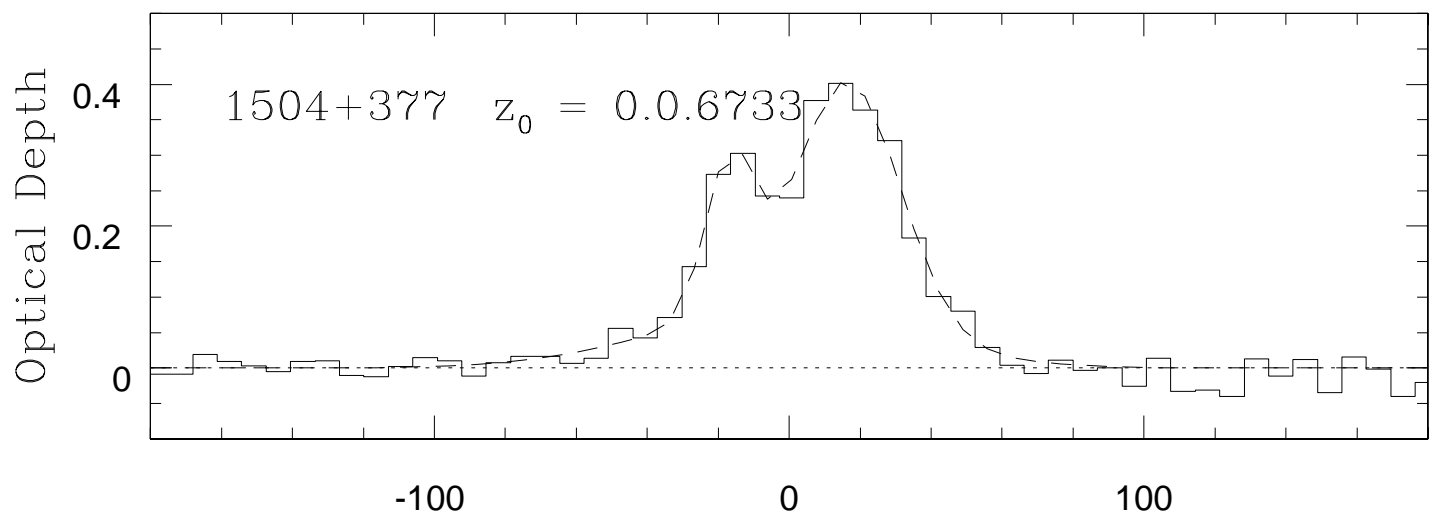
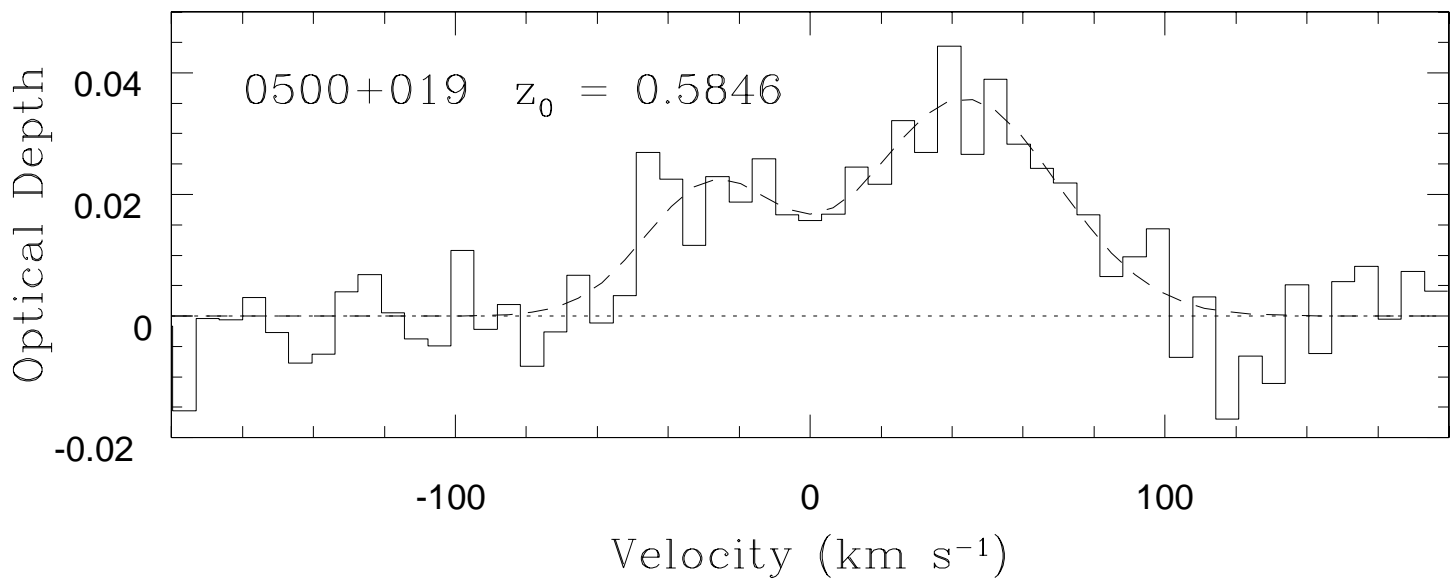
\end{document}